\documentstyle[twoside,fleqn,epsfig,espcrc2]{article}

\newcommand{\AmS}{{\protect\the\textfont2
  A\kern-.1667em\lower.5ex\hbox{M}\kern-.125emS}}

\title{The Alpha Magnetic Spectrometer (AMS): search for antimatter and dark matter on the 
International Space Station}
\author{R. Battiston\address{Dipartimento di Fisica e Sezione INFN, \\ 
        Via Pascoli, 06100 Perugia, Italy}}
       
\begin{document}

\begin{abstract}
The Alpha Magnetic Spectrometer (AMS) is a state of the art 
detector for the  extraterrestrial  study of antimatter, matter and missing matter. 
After a precursor flight on STS91 in may 1998, AMS will be installed 
on the International Space Station where it will operate for three years.
In this paper the AMS experiment is described and its  physics potential reviewed.

\end{abstract}

\maketitle

\section{ Introduction }

   The disappearance of cosmological antimatter and the pervasive presence of dark matter 
are two of  the greatest puzzles in the current  understanding of  the universe.

The Big Bang model  assumes that, at its  very beginning,  half  of the universe
 was made out  of antimatter. The validity of this model is based
on three key experimental observations: the recession of galaxies (Hubble expansion), 
the highly isotropic cosmic microwave background and  the relative abundances of
 light isotopes.  However, a fourth basic observation, the presence of cosmological 
 antimatter somewhere in the universe, is missing.  Indeed measurements of the 
intensity  of gamma ray flux in the MeV region exclude the presence of a significant
 amount of antimatter up to the scale of the local  supercluster 
of galaxies  (tens of Megaparsecs). It follows that, either  antimatter has been 
destroyed immediately  after the Big Bang by some  unknown mechanism, or matter and 
antimatter  were separated (by some other unknown mechanism) into different region
 of space, at scales larger than superclusters. 
 All efforts to reconcile the  
the absence of antimatter  with cosmological models 
 that do not require new physics failed 
(see \cite{Steigmann,Kolb,Peebles},
for a review of these theories). 

We are  then currently unable to explain the fate of half of the  
baryonic matter present at the beginning of our  universe.

Rotational velocities in spiral galaxies and dynamical effects in galactic clusters
provide us convincing evidence that, either Newton laws break down at scales of galaxies  
 or, more likely,  most (up to 99\%) of our universe consists of non-luminous  (dark) matter. 
There are several dark matter candidates.
They are commonly classified  as ''hot'' and ''cold'' dark matter, 
depending on their relativistic  properties at the time of decoupling from normal
 matter in   the early universe. As an example, light neutrinos are  
 obvious candidates for ''hot'' dark matter while  Weakly Interacting
 Massive Particles (WIMP's) are often considered as  ''cold'' dark matter candidates 
\cite{Ellis,Turner}. 

\begin{figure}[htb]
 \begin{center}
  \mbox{\epsfig{file=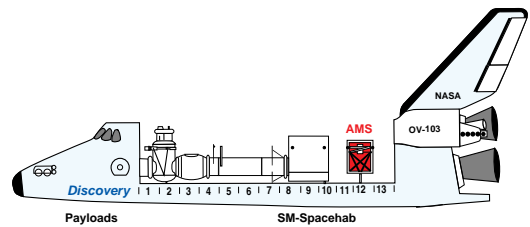,width=7 cm}}
  \caption{\em {AMS on STS 91  (Discovery)}}
 \end{center}
\end{figure}

\begin{figure*}[htb]
 \begin{center}
  \mbox{\epsfig{file=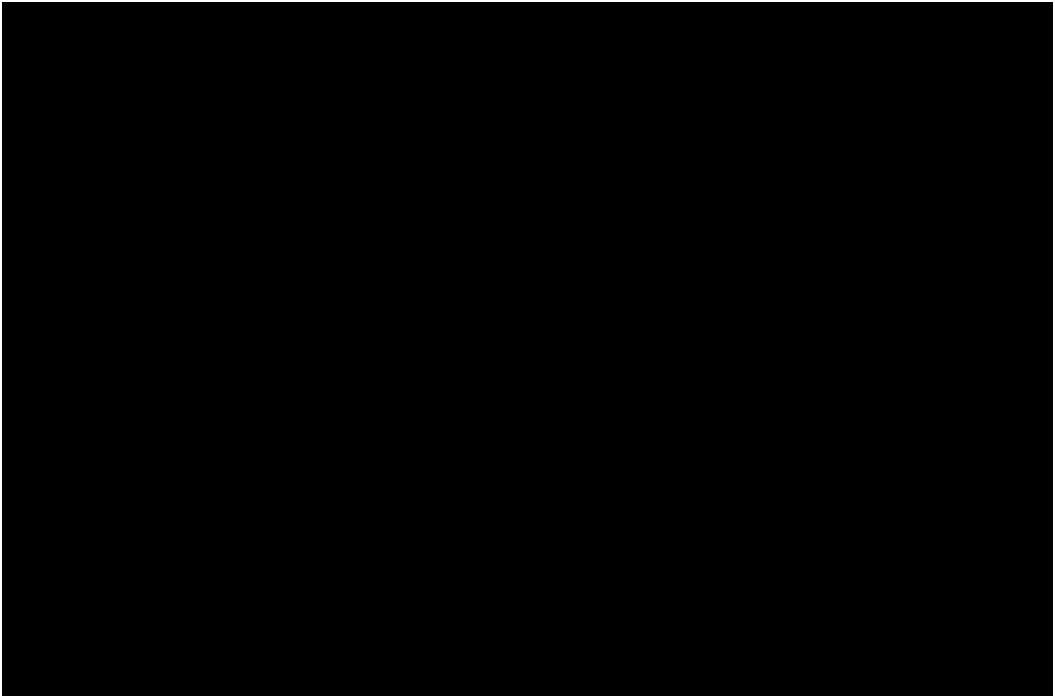,width=14cm}}
% \framebox[105mm]{\rule[-50mm]{0mm}{123mm}}
  \caption{\em {The  International Space Station Alpha; AMS will be installed 
on the left side of the main truss}}
 \end{center}
\end{figure*}

        In either cases we are currently unable to explain the origin of most
 of the mass of our universe.

To address these two fundamental questions in astroparticle physics
  a  state of the art detector, 
 the Alpha Magnetic Spectrometer (AMS) \cite{AMS}
has been recently approved  by NASA to operate on  the
 International Space Station Alpha (ISSA).

AMS  is manifested for a precursor flight with  STS91, (Discovery, may 1998, Figure 1), 
and for a three year long exposure on the International Space Station (ISS) (Figure 2),
 after its installation  during Utilization Flight n.4 
(Discovery, january 2002). 
AMS  has been proposed   and is being built by  an international collaboration involving  
China, Finland,  France, Germany, Italy, Rumenia,  Russia, Switzerland, Taiwan  and US. 

\begin{figure}[htb]
 \begin{center}
 \mbox{\epsfig{file=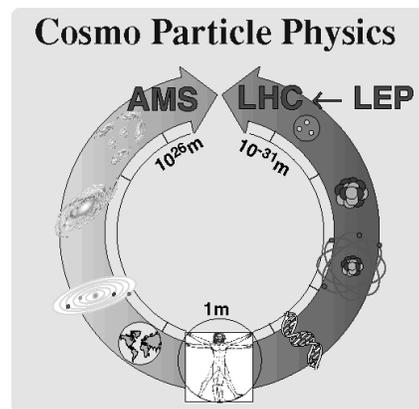,width=5.5cm}}
%\framebox[45mm]{\rule[-5mm]{0mm}{23mm}}
\vspace{0.1cm} 
\caption{\em {Astroparticle physics}}
\label{fig:astroparticle}
\end{center}
\end{figure}
\clearpage 

\cleardoublepage 

\section{ Particle physics and antimatter}

Particles  and antiparticles   are connected through  three discrete space-time 
symmetries  which are the foundations
 of relativistic field theory: 
charge conjugation (C), parity (P) and time reversal (T).  
We know today that each  of these 
symmetries can be independently  violated. 
 Invariance of fundamental interactions under the  three combined  transformations is, however, 
always maintained (CPT theorem). The best tests of the CPT theorem are indeed 
based on  the comparison of  lifetimes and masses  of  particle and antiparticles,
 since the a CPT tranformation links a particle to its antiparticle.
 An asymmetry between matter and antimatter in our universe
 would then be strictly related to a violation of  these discrete symmetries.
In 1967, Sakharov  noted that baryogenesis, which 
\begin{figure}[htb]
 \begin{center}
  \mbox{\epsfig{file=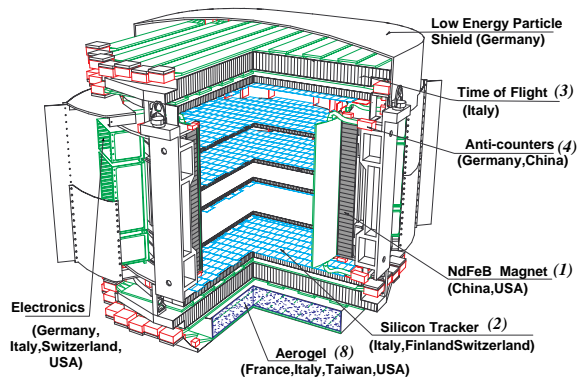,width=8cm}}
  \caption{\em {AMS configuration for the may 1998  Shuttle  percursor  flight (STS91)}}
 \end{center}
\end{figure}
describes the evolution from a symmetric universe to an asymmetric one, requires 
four ingredients, three of which, baryon number violation, C  and CP  violations
 are fundamental properties of elementary particles \cite{Sakharov}. Neither  particle 
acceleration experiments nor proton decay experiments have yet provided evidence for
baryon number violation. 
However, since there is a long range force associated 
with baryonic charge, there is no compelling reason for baryon number to be conserved.
Nevertheless, until particle  physics experimental data provides  confirmation of these
ideas, the observed  lack of antimatter in our part of universe will be the 
strongest evidence for baryon number violation.

CP violation has been observed directly  in three decays of the $K_L$ meson 
and nowhere else in particle phyisics experiments:

\begin{equation}
                        K_L \rightarrow \pi^+ \pi^-, \ \   
 K_L \rightarrow \pi^o \pi^o, \ \    K_L \rightarrow \pi l \nu  
\label{eq:s1}
\end{equation}

($l$ is either an electron or muon). The absence of antimatter in our part of the 
universe (and perhaps in the entire universe), combined with Sakharov conditions, 
provides the only other evidence for CP violation.

Since CP violation as the origin of baryogenesis occurs at energy scales much
greater and  at much greater level than in the   in the kaon system, the study of
baryon asymmetry in the  universe provides  crucial informations  in attempts to 
probe beyond the Standard Model of particle physics. 
Cosmological models that predict baryon symmetry on the scale of the observable 
universe may involve different sorts of CP violating mechanisms than those models which 
exclude antimatter altogether.  Then, the determination by AMS of whether the universe does 
or does not contain domains of antimatter beyond our local supercluster of galaxies 
is   of great importance to understand the  fundamental laws of  particle physics. 

        A comment about the study of CP violation on the B-meson system. The  B-factories currently
 being developed to study  CP violation 
in the B-meson system may provide information pertaining to baryon asymmetry.
However current understanding suggest that CP violation in the B-system is most likely 
unrelated to the CP  violation necessary to produce cosmological baryon asymmetry. 
In particular, just as CP violation,  present in the kaon system from a phase 
in the Cabibbo-Kobayashi-Maskawa (CKM) matrix, is too small to produce the observed 
cosmic asymmetry, similar conclusions are likely to hold for the B-system.
However it is possible that CP violation beyond the phase of the CKM matrix
 is operative in the B-system,  and   B-factories are likely to reveal 
whether or not this is the case. However, new sources of CP violation 
which have been suggested for the origin of the baryon asymmetry do not 
typically lead to significant new CP violation in the B-system.
AMS will then bring  informations on the origin of CP violation which are complementary  to 
those obtained at the accelerators (Figure 3). 

\begin{figure}[htb]
 \begin{center}
  \mbox{\epsfig{file=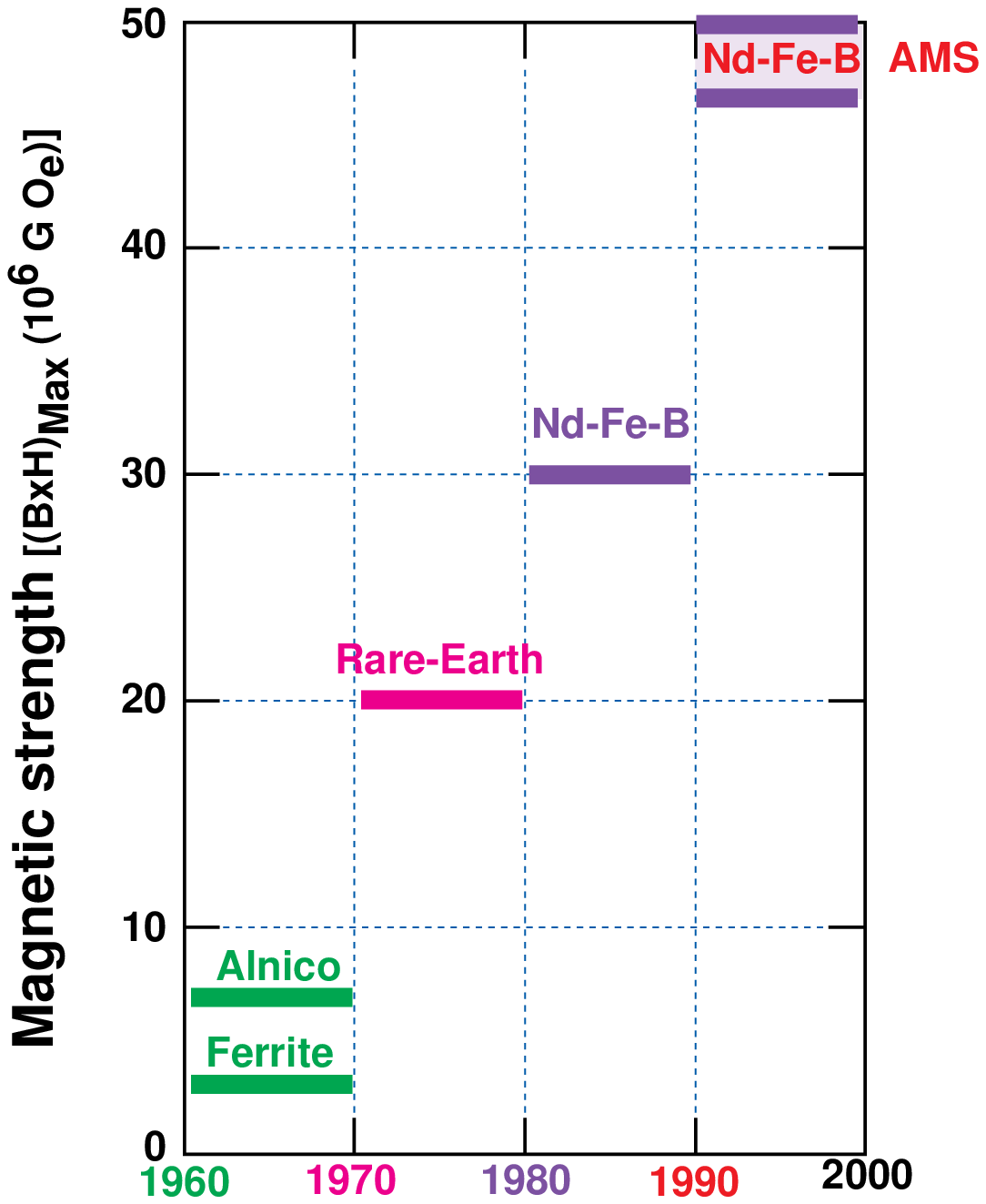,width=6.0cm}}
  \caption{\em {Progress on permanent magnets}}
 \end{center}
\end{figure}
\begin{figure}[htb]
 \begin{center}
  \mbox{\epsfig{file=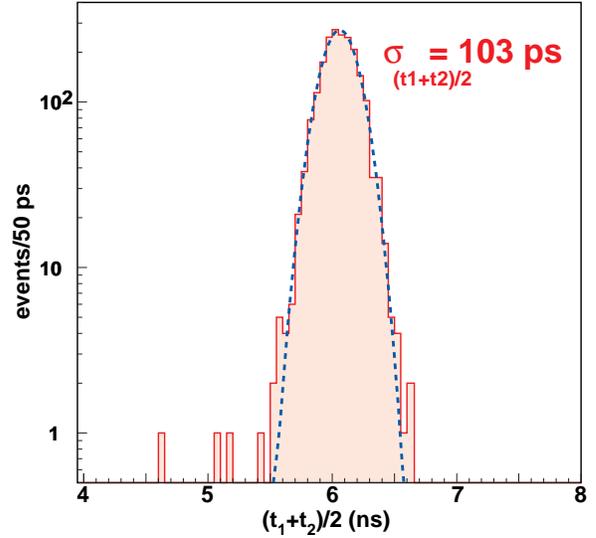,width=8.0cm}}
  \caption{\em {AMS Time of Flight detector  resolution as measured in a test beam}}
 \end{center}
\end{figure}

\begin{figure}[htb]
 \begin{center}
  \mbox{\epsfig{file=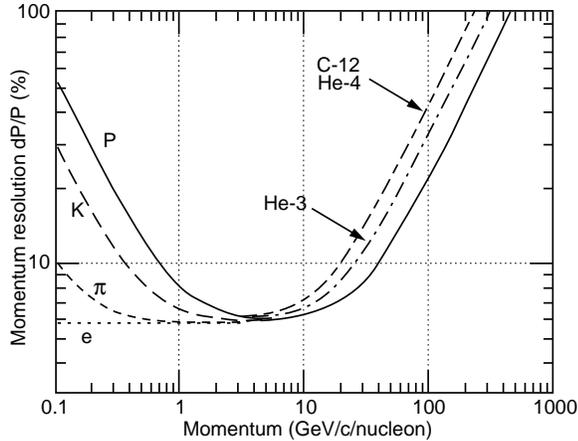,width=8.0cm}}
  \caption{\em { Momentum resolution of the  AMS  Silicon Tracker}}
 \end{center}
\end{figure}

\section{ AMS experiment design principles}
\par
  Search of antimatter requires the capability to
identify  with
 the highest degree of confidence, the  type of particle traversing 
the experiment together with the absolute value  and the sign of its electric charge.
This can be achieved through repeated 
measurements of the particle momentum (Spectrometer), velocity 
(Time of Flight, Cerenkov detectors)
and energy deposition (Ionization detectors).

        The experiment
configuration for the precurson flight is shown in Figure 4.
It consist of a large acceptance magnetic spectrometer (about $0.6\ m^2 sr$) made of 
 a new type  permanent  Nd-Fe-B magnet (1), surrounding   a six layer high
 precision silicon tracker (2)   and sandwiched
between four planes of the  Time of Flight scintillator system (ToF) (3). A scintillator anticounter system (4),
 located on the  magnet inner  wall
 and a  solid state Cherenkov (8)  detector, 
complete the experiment. 

The magnet is based on recent advancements in permanent magnetic material and technology (Figure 5)
 which make it possible to use very high grade Nd-Fe-B to construct 
a permanent magnet with $BL^2=\ 0.15\  Tm^2$ weighting $\le 2$ tons. 
A charged particle traversing the  spectrometer will  trigger the experiment  through the 
ToF  system. The ToF will measure the  particle  velocity with a resolution of
$\sim 100\ ps$ over a distance of $\sim \ 1.4\ m$ \cite{Palmonari} (Figure 6) .

 The momentum resolution
 of the  Silicon Spectrometer is given in Figure 7 \cite{Battiston}: at low rigidities, below 8 GV,
 its  resolution is dominated by multiple scattering  ($\frac{\Delta p}{p}\sim 7\% $)
 while the maximum detectable rigidity is about $500\ GV$. The parameters of the Silicon 
Spectrometer are given in Table 1.
Both the ToF scintillators and the silicon layers  measure   $\frac{dE}{dx}$, allowing a 
multiple determination of the absolute value of the particle charge.  

\begin{table}[thb]
  \caption{AMS silicon tracker parameters}
 \begin{center}
  \begin{tabular}{|c|c|} \hline
   Number of planes & $6$  \\ 
Accuracy (bending plane) & $10\  \mu m$  \\
Accuracy (non bending plane) & $30\  \mu m$  \\
Number of channels & 172000   \\
Power consumption & $400 \ W$ \\  
Weight & $130 \  kg$   \\
Silicon Area (double sided) & $6 \ m^2$   \\ \hline
  \end{tabular}
 \end{center} 
\end{table}

\section{ AMS  physics potential}

        The physics objectives of AMS are:

\begin{itemize}
\item   search for Antimatter ($\bar{He}$ and $\bar{C}$) in space with a sensitivity 
of $10^4$ to $10^5$ better than current limits.

\end{itemize}
 The breakdown of the Time-Reversal 
symmetry in the early universe might have set different sign for the production 
of matter and antimatter in different regions of space. Since there are $O(10^8)$ 
super clusters of galaxies and the observational constraints are limited to the scale 
of the local supercluster, the universe can be symmetric on a larger scale. 
The observed matter-antimatter asymmetry would then be  a local phenomenon.
 Figure 8a,b shows the sensitivity of AMS, after three years
on ISSA, in detecting 
$\bar{He}$ and $\bar{C}$ compared to the current limits.  In Figure 8a we have included the 
prediction of $\bar{He}$ yield as predicted by a model assuming  a matter-antimatter
symmetric universe at the level of super-cluster of galaxies \cite{Ahlen}. 
As seen from the figure,  
the current limits are  not sensitive enough to test the existence of superclusters
 of galaxies
of antimatter and AMS is $10^3$ times more sensitive than  this prediction
 based on matter-antimatter  symmetric universe.
        
\begin{figure*}[htb]
 \begin{center}
  \mbox{\epsfig{file=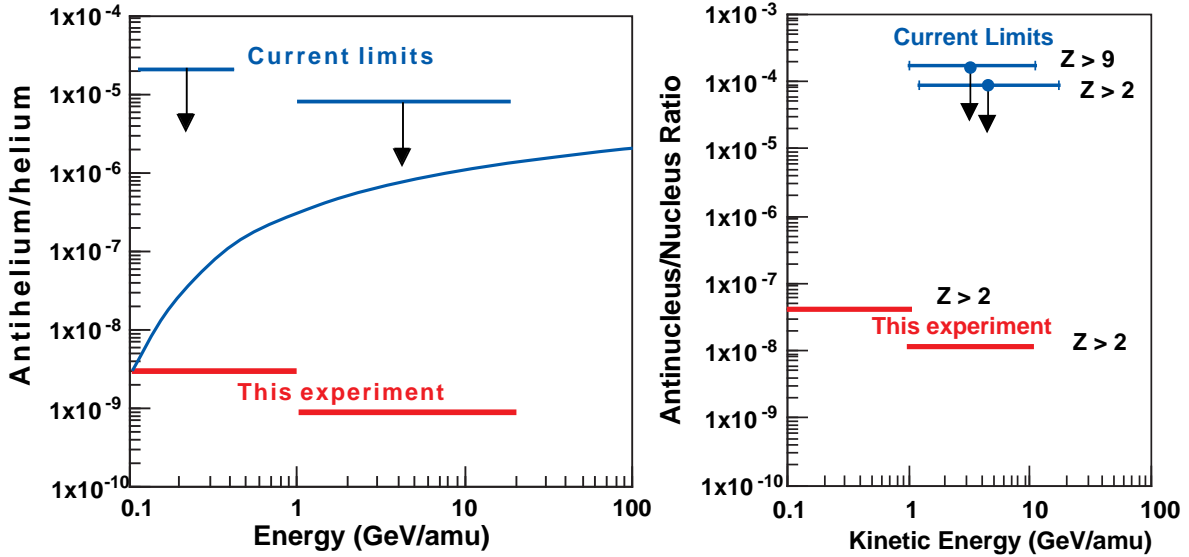,width=16cm}}
  \caption{\em { Sensitivity of AMS (3 years on ISSA) in a search for  (a)  $\bar{He}$; (b)
   $Z>2$  antinuclei ($95\%$ C.L.).  $\bar{He}$  sensitivity is  compared 
to a prediction assuming a matter-antimatter symmetric universe at the 
level of super-cluster of galaxies } }
 \end{center}
\end{figure*}

\begin{figure}[htb]
 \begin{center}
  \mbox{\epsfig{file=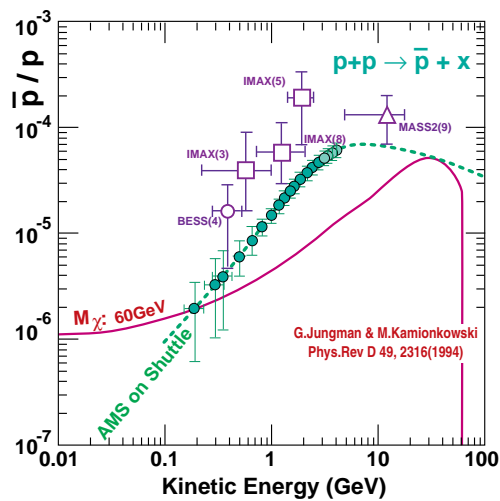,width=7cm}}
  \caption{\em {Simulated 100 hour shuttle flight $\bar{p}$ measurement by AMS, compared with 
a compilation of existing $\bar{p}$ data}}
 \end{center}
\end{figure}

\begin{figure}[htb]
 \begin{center}
  \mbox{\epsfig{file=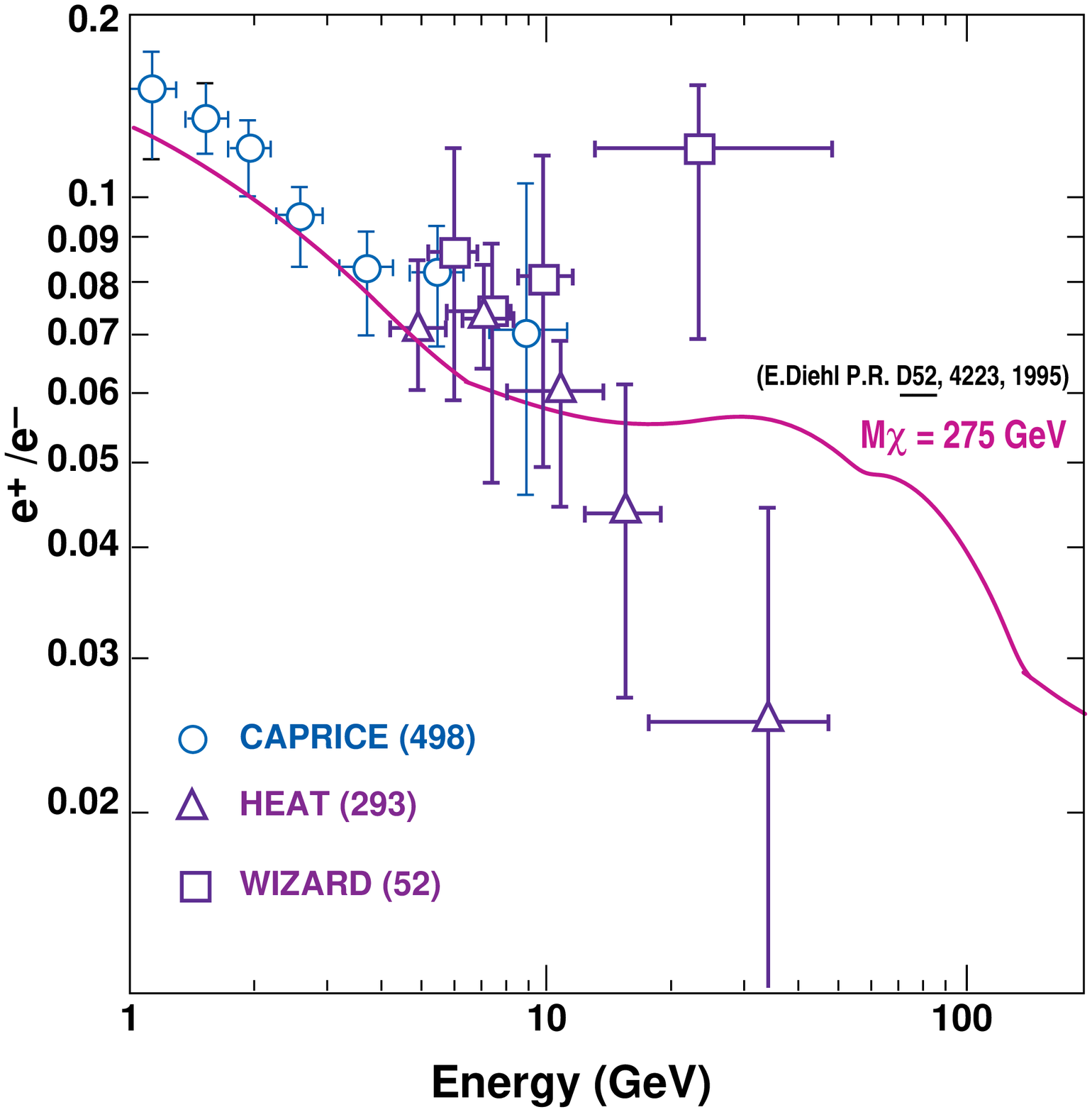,width=7cm}}
  \caption{\em { Simulated 3 years  AMS  $e^+$ measurement compared with a 
compilation of existing $e^+$ data}}
 \end{center}
\end{figure}

\begin{itemize}
\item  search for dark matter by high statistics precision measurements of   
  $\bar{p}$, $e^+$  and $\gamma$ spectra.
\end{itemize}
 These measurements will allow direct searches 
for the various annihilation and decay products of WIMP's in the galactic halo:

\begin{equation}
                        \bar{\chi} + \chi \rightarrow \bar{p}+X, e^+ + X, 2\gamma
 \end{equation}

\begin{equation}
 \chi,\bar{\chi} \rightarrow \gamma  \nu\  
\end{equation}

        Figure 9 shows the simulation of the 100 hour measurement of the $\bar{p}$ flux 
 on the precursor flight compared to a compilation of  existing $\bar{p}$ data together 
with a prediction of the effects of an heavy neutralino  \cite{Jungman}.

 Similarly Figure 10 shows the simulation of three years
of AMS $e^+$ data compared to the existing measurements. The accuracy of the  AMS measurement
of the $\frac{e^+}{e^-}$ spectrum would allow  to  test the existence of an heavy neutralino
($m_\chi\sim100\ GeV$) \cite{Diehl}.
Also the measurement of the $\gamma$ spectrum will test the existence of 
neutralinos in case of  R-parity violating SUSY models, through 
its decay in two particles  with $E_{\gamma,\nu}=m_\chi/2$ 
\cite{Bergstrom,Berezinsky,Stecker}.

\begin{table*}[hbt]
% space before first and after last column: 1.5pc
% space between columns: 3.0pc (twice the above)
\setlength{\tabcolsep}{1.5pc}
% -----------------------------------------------------
% adapted from TeX book, p. 241
\newlength{\digitwidth} \settowidth{\digitwidth}{\rm 0}
\catcode`?=\active \def?{\kern\digitwidth}
% -----------------------------------------------------
\caption{Physics capabilities of AMS after  three years on ISSA}
\begin{tabular*}{\textwidth}{@{}l@{\extracolsep{\fill}}rrrr} \hline
% \begin{tabular}{|c|c|c|c|c|} \hline
   Elements & Yield/sensitivity & (Now) & Range (GV) & Physics  \\ \hline
$e^+$ & $10^8$ & $(\sim 10^3)$ & $0.1-100$   & Dark Matter    \\
$\bar{p}$ & $500000$ & 
$(\sim 30)$ & $0.5-100$   & SUSY    \\
$\gamma$ &  &  & $0.1-300$   &  R-parity  \\ \hline
$\bar{He}$/$He$  & $\frac{1}{10^9}$ & $(\frac{1}{10^5})$ & $0.5-20$ 
 & Antimatter \\  
$\bar{C}$/$C$  & $\frac{1}{10^8}$ & $(\frac{1} { 10^4})$ & $0.5-20$ 
 & CP vs GUT , EW   \\ \hline
$D, H_2$ & $10^9$ & $ $ & $1.0-3.0$   & Astrophysics  \\
$\ ^3He$/$\ ^4He$ & $10^9$ & $ $ & $1.0-3.0$   &  CR propagation \\
$\ ^{10}Be$/$\ ^9Be$ &$2\%$  &  & $1.0-3.0$   & CR confinement   \\ \hline
  \end{tabular*}
\end{table*}

\begin{itemize}
\item   astrophysical studies  by high statistics precision
 measurements of $D,\ ^3He, B, C,$ $ \ ^9Be$ and $\ ^{10} Be$ spectra.  
\end{itemize}

         Precision measurement of isotopes and elemental abundances  in Cosmic Rays (CR) 
give very important informations 
about CR   origin, their   galactic confinement time and
 their  propagation inside and between galaxies. This is deeply related
to the search for  antinuclei, since antimatter particles, to be detected, should excape a region of the 
universe dominated by antimatter, propagate through the intergalactic void separating
superclusters and enter our galaxy. As an example AMS will measure 
in three months $5\ 10^7$  deuterium events reducing the current uncertainty  on  $D/p$ 
by a factor of 100. Similarly, AMS will be able to separe  $\ ^3He$ from $\ ^4He$ up to 
$5\ GV$ of rigidity, thus  collecting in three years  $\sim 4\ 10^8$ $\ ^3He$ 
and $\sim 4\ 10^9$ $\ ^4He$,  reducing the existing uncertainties by a factor 200.
The ratio $\frac{B}{C}$, which constrains the parameters regulating the outflowing 
galactic wind, can be measured by AMS better than the current data within one day   and 
up to 100 GV of rigidity. On the other side, the ratio  $\frac{^{10}Be}{^9Be}$ determines
 the CR confinement time; at present, a dozen of $\ ^{10}Be$ events  have been detected during  more than 
forteen years  of observations and the CR galactic confinement time is 
currently known within  a large uncertainty ($\sim 40\%$).
  With AMS it will be possible to detect tens of 
$\ ^{10}Be$/day in the $1\  GV$ range, thus dramatically reducing the error on  CR
 confinement time.

        In addition, the capability  of AMS of detecting 
high energy $\gamma$ rays would  make it possible to perform  very important 
 observations  in gamma ray astrophysics  after the turn off of the EGRET experiment 
(within one or two years).
Simulations suggest indeed that the performances of  AMS in monitoring the $\gamma$-sky 
 in the multi GeV range
are comparable to these of   EGRET (however AMS  on the Space Station will not be able
 to point to targets of opportunity)   \cite{Fiandrini,Battiston2}.
 A comparison between the two experiments is shown in 
Table 3.

\begin{table*}[hbt]
% space before first and after last column: 1.5pc
% space between columns: 3.0pc (twice the above)
\setlength{\tabcolsep}{1.5pc}
% -----------------------------------------------------
% adapted from TeX book, p. 241
%\newlength{\digitwidth} \settowidth{\digitwidth}{\rm 0}
%\catcode`?=\active \def?{\kern\digitwidth}
% -----------------------------------------------------
\caption{Comparison of EGRET and AMS $\gamma$ detection capabilities}
\begin{tabular*}{\textwidth}{@{}l@{\extracolsep{\fill}}rrrr} \hline
 % \begin{tabular*}{|c|c|c|} \hline
  Parameter  &EGRET & AMS  \\ \hline
Peak effective area ($cm^2sr$) & $1500$ & $900$   \\
Angular resolution $(68\%)$  & $1.7^o(\frac{E_\gamma}{1\ GeV})^{-0.534}$ & 
$1.27^o(\frac{E_\gamma}{1\ GeV})^{-0.874}$  \\
Mean opening angle & $\sim 25^o$  & $\sim 30^o$     \\  
Total viewing time & $\sim 2.5\ yr$ & $\sim 3\ yr$ \\  
Source flux sensitivity ($>1 GeV$) & $\sim 10^{-8} cm^{-2}s^{-1}$& 
$\sim 10^{-8} cm^{-2}s^{-1}$ \\  
Detector energy range  ($GeV$) & $0.02\ to \sim 20$ & $0.3\ to \sim 200 $      \\ \hline
  \end{tabular*}
\end{table*}

 Before the
advent of new generation gamma ray space facilities like GLAST  \cite{Bloom}, AMS will then be the only space
born experiment to continue the EGRET mission  at the beginning of the next century, 
covering the interesting region $E_\gamma=30 - 300 GeV$. These data will extend our 
knowledge in key areas of $\gamma$ astrophysics like:  Active Galactic Nuclei (AGN), blazars,
 $\gamma$-bursters. They will also allow  a measurement of the extragalactic  $\gamma$ background
 and of the  interaction of high energy $\gamma$  rays with the 
 black-body microwave background.

        The physics capabilities of AMS after three years of exposure on the  ISSA are summarized in 
Table 2.

\section {Conclusion}

 During the past forty years, there have been many fondamental discoveries in astrophysics
measuring $UV$, $X$-ray and $\gamma$-ray photons. There has never been a sensitive magnetic 
spectrometer in space, due to the extreme difficulty and very high cost of putting 
a superconducting magnet in orbit. AMS will be the first large acceptance magnetic 
detector in space.  It will allow a measurements  of the flux of all kind of  cosmic
rays with an accuracy orders of magnitude better than before. The
large improvement in sensitivity given by this new instrument, will  allow us to 
 enter into a totally new domain to explore the unknown.

\end{document}